\documentclass{article}
\usepackage{spconf,amsmath,graphicx}

\usepackage{multirow}
\usepackage{hyperref}
\usepackage{booktabs}

\usepackage[noend]{algpseudocode}
\usepackage{algorithmicx,algorithm}
\usepackage{setspace}
\usepackage{amssymb,amsmath,bm}
\usepackage{subcaption}

\newcommand{\mybff}[1]{\bm{#1}}

\title{Incremental FastPitch: Chunk-based High Quality Text to Speech}

\name{Muyang Du$^1$, Chuan Liu$^1$, Junjie Lai$^1$}
\address{$^1$NVIDIA Corporation}

\begin{document}

\maketitle

\begin{abstract}
Parallel text-to-speech models have been widely applied for real-time speech synthesis, and they offer more controllability and a much faster synthesis process compared with conventional auto-regressive models. Although parallel models have benefits in many aspects, they become naturally unfit for incremental synthesis due to their fully parallel architecture such as transformer. In this work, we propose Incremental FastPitch, a novel FastPitch variant capable of incrementally producing high-quality Mel chunks by improving the architecture with chunk-based FFT blocks, training with receptive-field constrained chunk attention masks, and inference with fixed size past model states. Experimental results show that our proposal can produce speech quality comparable to the parallel FastPitch, with a significant lower latency that allows even lower response time for real-time speech applications.
\end{abstract}
\section{Introduction}
\label{sec:intro}

In recent years, Text-to-Speech (TTS) technology has witnessed remarkable advancements, enabling the generation of natural and expressive speech from text inputs. Neural TTS system primarily contains an acoustic model and a vocoder. It involves first converting the texts to Mel-spectrogram by acoustic models such as Tacotron 2\cite{shen2018natural}, FastSpeech\cite{ren2019fastspeech}, FastPitch\cite{lancucki2021fastpitch}, GlowTTS\cite{kim2020glow}, then converting the Mel feature to waveform by vocoders such as WaveNet\cite{van2017wavenet}, WaveRNN\cite{kalchbrenner2018efficient, du23_interspeech}, WaveGlow\cite{prenger2019waveglow}, and HiF-GAN\cite{kong2020hifi}. Moreover, with the boost of real-time and streaming applications, there is an increasing demand for TTS systems capable of producing speech incrementally, also known as streaming TTS, to provide lower response latency for better user experience. For example, Samsung\cite{ellinas2021high} proposed a low-latency streaming TTS system running on CPUs based on Tacotron 2 and LPCNet\cite{valin2019lpcnet}. NVIDIA\cite{du2023efficient} also proposed a highly efficient streaming TTS pipeline running on GPUs based on BERT\cite{kenton2019bert}, Tacotron 2 and HiFi-GAN. Both of them uses auto-regressive acoustic model for incremental Mel generation.

Auto-regressive acoustic models such as Tacotron 2 is capable of producing natural speech by leveraging sequential generation to capture prosody and contextual dependencies. However, it suffers from slow inference due to the frame-by-frame generation process and susceptibility to over-generation and word-repeating artifacts due to unstable alignment learned between the input phonemes and output frames. In contrast, parallel acoustic models like such as FastPitch offers a faster inference process by producing complete Mel-spectrogram in one step. Additionally, it also shows benefits in providing the flexibility to manipulate pitch, duration, and speed of the synthetic speech as those metadata are pre-generated before decoding.

\begin{figure}[t]
  \centering
  \includegraphics[width=\linewidth]{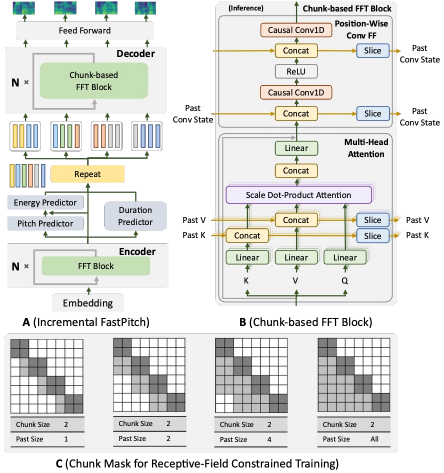}
  \caption{Incremental FastPitch, Chunk-based FFT Block, and Chunk Mask for Receptive-Filed Constrained Training}
  \label{fig:inference_timeline}
\vspace{-1em}
\end{figure}

Although parallel acoustic models offer many advantages, their model structure poses challenges for their use in incremental speech synthesis. For instance, FastPitch utilizes a transformer\cite{vaswani2017attention} decoder, wherein attention is computed across the entire encoded feature sequence to generate the Mel-spectrogram output. A straightforward method is to slice the encoded feature sequence into chunks and then decode each chunk into a corresponding Mel chunk. However, this approach compels the decoder to focus only on a chunk, resulting in audible discontinuity at the edges of Mel chunks, even when overlapping between chunks is used. An alternative approach is to modify the model to use an auto-regressive decoder. However, this fails back to frame-by-frame generation, sacrificing the parallelism advantage. Therefore, an ideal decoder for incremental TTS should be able to incrementally generate Mel chunks while maintaining parallelism during the chunk generation process and keeping the computational complexity of each chunk consistent in the meantime.

Based on the above considerations, we present Incremental FastPitch, capable of producing high-quality Mel chunks while maintaining chunk generation parallelism and providing low response latency. We incorporate chunk-based FFT blocks with fixed-size attention state caching, which is crucial for transformer-based incremental TTS to avoid the computational complexity increases with synthetic length. We also utilize receptive-filed constrained training and investigate both the static and dynamic chunk masks, which is vital to align the model with limited receptive-filed inference. 

\section{Method}

\subsection{Incremental FastPitch}

Figure 1A depicts the proposed Incremental FastPitch model, a variant of the parallel FastPitch. It takes a complete phoneme sequence as input and generates Mel-spectrogram incrementally, chunk-by-chunk, with each chunk contains a fixed number of Mel frames. Incremental FastPitch is equipped with the same encoder, energy predictor, pitch predictor, and duration predictor as the parallel FastPitch. However, the decoder of Incremental FastPitch is composed of a stack of chunk-based FFT blocks. In contrast to the decoder of parallel FastPitch that takes the entire upsampled unified feature $\mybff{\Bar{u}}$ as input and generate the entire Mel-spectrogram at once, The decoder of Incremental FastPitch first divide the $\mybff{\Bar{u}}$ to $N$ chunks $[\mybff{\Bar{u}_{1}}, \mybff{\Bar{u}_{2}}, ..., \mybff{\Bar{u}_{N}}]$, then convert one chunk $\mybff{\Bar{u}_{i}}$ at a time to a chunk of Mel $\mybff{\Bar{y}_{i}}$. During training, we apply a chunk-based attention mask on the decoder to help it adjust to the constrained receptive field in incremental inference, which we term it as the Receptive Field-Constrained Training.

\subsection{Chunk-based FFT Block}

Figure 1B illustrates the chunk-based FFT block, which contains a stack of a multi-head attention (MHA) block and a position-wise causal convolutional feed forward block. Compare with parallel FastPitch, the MHA block in the chunk-based FFT block requires two additional inputs: past key and past value, produced by itself during previous chunk generation. Instead of utilizing all the accumulated historical past keys and values from prior chunks, we employ fixed-size past key and value for inference by retaining only their tails. The past size maintains consistent throughout incremental generation, preventing an increase in computational complexity with the number of chunks. Although we impose an explicit past size limit, experiments shows that it is capable of encoding sufficient historical information for generating high-quality Mel. The calculation of MHA is defined as:

\begin{equation}
\begin{aligned}
k^{t}_{i} &= \mathrm{concat}(pk^{t-1}_{i}, KW^{K}_{i}) \\
v^{t}_{i} &= \mathrm{concat}(pv^{t-1}_{i}, VW^{V}_{i}) \\
o^{t}_{i} &= \mathrm{attention}(k^{t}_{i}, v^{t}, QW^{Q}_{i}) \\
o^{t}_{M} &= \mathrm{concat}(o^{t}_{1}, ..., o^{t}_{h})W^{O} \\
pk^{t}_{i} &= \mathrm{tail\_slice}(k^{t}_{i}, S_{p}) \\
pv^{t}_{i} &= \mathrm{tail\_slice}(v^{t}_{i}, S_{p}) 
\end{aligned}
\end{equation}

where $pk^{t-1}_{i}$ and $pv^{t-1}_{i}$ are the past $K$ and past $V$ of head $i$ from chunk $t-1$. $k^{t}_{i}$ and $v^{t}_{i}$ are the embedded $K$ and $V$ with the past concatenated along the time dimension for attention computation of head $i$ at chunk $t$. $o^{t}_{M}$ is the output of MHA block at chunk $t$. $W^{K}_{i}$, $W^{V}_{i}$, $W^{Q}_{i}$, and $W^{O}$ are the trainable weights. $S_{p}$ is the configurable fixed size of the past. $pk^{t}_{i}$ and $pv^{t}_{i}$ are obtained by slicing size $S_{p}$ from the tail of $k^{t}_{i}$ and $v^{t}_{i}$ along the time dimension.

Similarly, the calculation of position-wise causal convolution feed forward block is defined as:
\begin{equation}
\begin{aligned}
c^{t}_{1} &= \mathrm{concat}(pc^{t-1}_{1}, o^{t}_{M}) \\
o^{t}_{c_{1}}  &= \mathrm{relu}(\mathrm{causal\_conv}(c^{t}_{1})) \\
c^{t}_{2} &= \mathrm{concat}(pc^{t-1}_{2}, o^{t}_{c_{1}}) \\
o^{t}_{c_{2}}  &= \mathrm{relu}(\mathrm{causal\_conv}(c^{t}_{2})) \\
pc^{t}_{1} &= \mathrm{tail\_slice}(c^{t}_{1}, S_{c_{1}}) \\
pc^{t}_{2} &= \mathrm{tail\_slice}(c^{t}_{2}, S_{c_{2}}) 
\end{aligned}
\end{equation}

where $pc^{t-1}_{1}$ and $pc^{t-1}_{2}$ are the past states of the two causal convolutional layers. Starting with $pc^{t-1}_{1}$, it's concatenated with $o^{t}_{M}$ to yield $c^{t}_{1}$, serving as input for the first causal conv layer. Next, $o^{t}_{c_{1}}$, the output from the first causal conv layer, is concatenated with $pc^{t-1}_{2}$ to generate $c^{t}_{2}$. This is then input to the second causal conv layer, resulting in the final output $o^{t}_{c{2}}$. Lastly, $pc^{t}_{1}$ and $pc^{t}_{2}$ are extracted by slicing sizes $S_{c_{1}}$ and $S_{c_{2}}$ from the tail of $c^{t}_{1}$ and $pc^{t}_{2}$ along the time dimension, respectively. Unlike the configurable $S_{p}$, we set $S_{c_{1}}$ and $S_{c_{2}}$ to their respective conv kernel sizes minus 1, which is adequate to attain equivalence with parallel inference.

\subsection{Decoder Receptive Field Analysis}

Figure 2 demonstrates the receptive filed of the proposed chunk-based decoder. For better visualization, we omit the positional-wise convolutional feed-forward blocks. The orange block at the top-right corner represents the final FFT output $O_t$ of chunk $t$. The dark green MHA blocks are those whose multi-head attention, past key, and past value outputs contribute to $O_t$. The light green MHA blocks are those whose past key and past value outputs contribute to $O_t$. Similarly, the blue blocks (past keys and past values) and the yellow blocks (inputs of green MHA blocks) are those who contribute to $O_t$. By feeding the fixed size past key and past value of chunk $t-1$ to each MHA block during chunk $t$ generation, we can expand the receptive field of chunk $t$ to several of its previous chunks without the need to explicitly feed those previous chunks as decoder input. 

The receptive field $\mathcal{R}$ depends on the number of decoder layers and the size of past keys and past values, as given by:

\begin{equation}
\begin{aligned}
\mathcal{R} = (N_{d} + \lfloor S_{p} / S_{c} \rfloor + 1) \cdot S_{c}
\end{aligned}
\end{equation}

where $N_{d}$ is the number of decoder layers, $S_{p}$ is the size of past keys and past values, and $S_{c}$ is the size of the chunk. The unit of $\mathcal{R}$ is the number of decoder frames. If $S_{p}$ is less than or equal to $S_{c}$, then the past key and past value output by a MHA block only depends on the input of that MHA block, thus $\mathcal{R}$ simply equals to $(N_{d}+ 1) \cdot S_{c}$, same as shown in figure 2, whereas if $S_{p}$ is greater than $S_{c}$, then the past key and past value of a MHA block at chunk $t$ will also depends on the past keys and values of that MHA block at previous chunks, resulting in $\mathcal{R}$ grows linearly with the floor of $S_{p}/S_{c}$.

\begin{figure}[t]
  \centering
  \includegraphics[width=\linewidth]{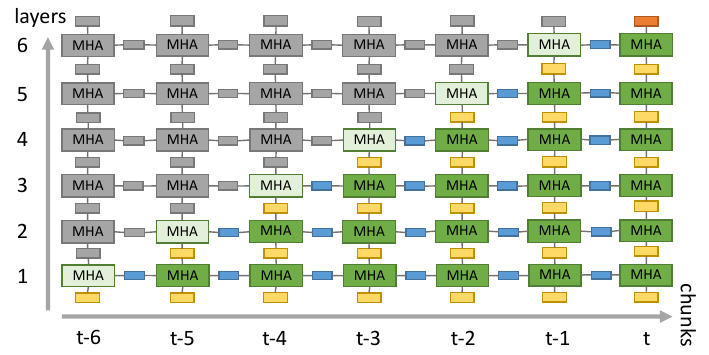}
  \caption{Chunk-based decoder receptive field visualization.}
  \label{fig:real_time_factor}
\vspace{-1.5em}
\end{figure}

\subsection{Receptive Field-Constrained Training}

Given a limited decoder receptive field during inference, it becomes vital to align the decoder with this constraint during training. Therefore, we use the Receptive Field-Constrained Training by applying chunk-based attention mask to all the decoder layers. Figure 1C visualizes various attention masks with a given chunk size (dark grey) and different past sizes (light grey). An intuitive approach is to randomly select a chunk size and past size for dynamic mask creation for each text-audio training data pair within a batch. This approach is similar to the masks used in the WeNet\cite{yao2021wenet,zhang2022wenet} ASR encoder. The dynamic mask can help the decoder generalize to diverse chunk and past sizes. However, most of the incremental system TTS employs a fixed chunk size for inference. Using a dynamic mask for training may potentially introduce a gap between training and inference. Therefore, we also investigate training with static masks that constructed using a fixed chunk size and past size during the training process.

\section{Experiments}

\subsection{Experimental Setup}

\noindent\textbf{Dataset.} The Chinese Standard Mandarin Speech Corpus\cite{csmscdatabaker} released by DataBaker is used for both training and evaluation. It contains 10,000 48kHz 16bit audio clips of a single Mandarin female speaker and has a total of 12 hours with each audio clip contains a short sentence of 4.27 seconds on average. In our experiments, we downsample the corpus to 22.05kHz and 100 audio clips are reserved for evaluation.

\noindent\textbf{Model \& Acoustic Specifications.} The proposed model parameters follow the open-source FastPitch implementation\cite{fastpitchcode}, except that we use causal convolution in the position-wise feed forward layers. The decoder is used to predict Mel-spectrogram with 80 frequency bins. It is generated through an FFT size of 1024, a hop length of 256 and a window length of 1024, applied to the normalized waveform. To enhance convergence speed and stability, the Mel values are standardized within a symmetrical range from -4 to 4.

\noindent\textbf{Training \& Evaluation.} 
Our models are trained using the Adam optimizer\cite{kingma2014adam} with batch size 8, initializing with a learning rate of 1e-4 and a weight decay of 1e-6. The experiments are performed on an NVIDIA RTX 6000 GPU, utilizing single precision and applying gradient clipping\cite{chen2020understanding}. We use Mel-spectrogram distance (MSD) and mean opinion score (MOS) to measure the speech quality. To ensure the Mel-spectrograms of two audios are properly aligned for MSD calculation, we first use a trained parallel FastPitch to produce unified duration, pitch, and energy values for evaluation texts, then use these values to process the output feature of Incremental FastPitch encoder. Regarding the MOS, we synthesize waveform for evaluation with HiFi-GAN trained using the same dataset as FastPitch. Since we focus on optimizing acoustic model for incremental TTS, the vocoding process is non-incremental. For Incremental FastPitch, we concatenate all the Mel chunks to the complete Mel for vocoding. The MOS scores are collected through the assessment of 20 evaluation samples for each configuration by 10 Amazon MTurk listeners, who assign scores ranging from 1 to 5. For audio samples, please refer to GitHub page\footnotemark[1].

\footnotetext[1]{\href{https://muyangdu.github.io/incremental-fastpitch}{\footnotesize{https://muyangdu.github.io/incremental-fastpitch}}}

\subsection{Discussion}
\begin{figure}[t]
  \centering
  \includegraphics[width=\linewidth]{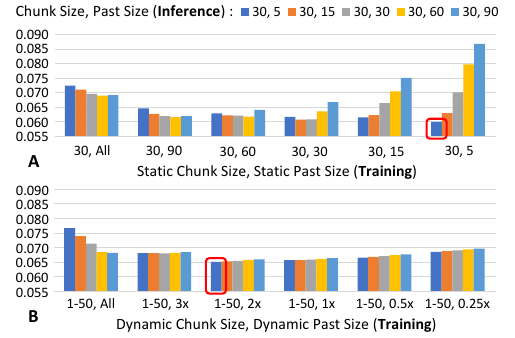}
  \caption{MSD between the parallel FastPitch and the Incremental FastPitch trained with different types of masks, then inference with different chunk and past sizes. Each bar in the figure represents a specific (chunk size, past size) for \textbf{inference}. The horizontal axis describes the (chunk size, past size) used for \textbf{training}. \textbf{A}. Static Mask. \textbf{B}. Dynamic Mask. }
  \label{fig:real_time_factor}
 \vspace{-1.5em}
\end{figure}

\subsubsection{Comparison of Static and Dynamic Chunk Masks}
Figure 3 shows the Mel-spectrogram distance between the Incremental FastPitch and the parallel FastPitch. During inference, we use a fixed chunk size 30 for all the models. In the sub-figure A, the models are train with static chunk masks. The chunk sizes are fixed to 30 and past sizes are set to 0, 5, 15, 30, 60, 90, and all. We can observe that the smallest MSD of each model is often achieved when we use the same (or similar) chunk size and past size for training and inference. The smallest MSD is achieved with past size 5 (red marked). Specifically, we find that if the model is trained with a small past size such as 5, it has a high MSD when inference with a big past size such as 90. On the contrary, if the model is trained with a big past size, it has a more stable MSD when inference with small past sizes. This observation suggests that even if the model is trained with a larger past context, it still learns to generate Mel chunk condition on nearby past contexts, rather than those far from the current chunk.

In the sub-figure B, the models are trained with dynamic chunk masks. The chunk sizes are randomly selected from range 1 to 50, and the past sizes are set to 0, 0.25, 0.5, 1, 2, 3 times of the selected chunk size and all. We observe that the MSD are more stable and similar if the inference past size changes, compared with static mask. The smallest MSD is achieved when we use 2 times of the randomly selected chunk size as the past size. However, the MSD of the dynamic chunk mask models are generally higher than the static chunk mask models. This observation confirms our suspicion raised in subsection 2.4 that dynamic mask training can introduce a training inference mismatch. Based on the above analysis, it is suggested to use a static mask for the best quality if the inference chunk and past sizes can be known in advance.

\subsubsection{Visualized Ablation Study}
We perform visualized ablation study to investigate the necessity of using past key value and past conv state. Figure 4 shows the synthetic Mel-spectrograms of parallel FastPitch and Incremental FastPitch. We can observe that the Incremental FastPitch can generate Mel with almost no observable difference compared with parallel FastPitch. However, if either the past key value or the conv state is removed, apparent discontinuation can be found between adjacent Mel chunks.

\begin{figure}[t]
  \centering
  \includegraphics[width=\linewidth]{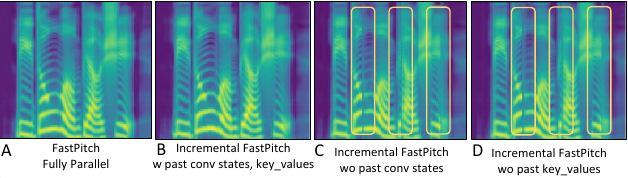}
  \caption{Mel-spectrogram Visualization.}
  \label{fig:real_time_factor}
\vspace{-1.5em}
\end{figure}

\subsubsection{Evaluation of Speech Quality and Performance}

To study the audible speech quality of both the static (S) and dynamic (D) mask trained Incremental FastPitch, we perform listening tests on the best S and D models selected based on the MSD analysis (marked as red in figure 3). As shown in table 1, we find that Incremental FastPitch is capable of producing high quality speech comparable with the parallel FastPitch. Furthermore, the score of D model is only slightly lower than the S model, although the D model has a 8.3\% higher MSD compared with the S model. This result shows that the audible difference of the S and D model is barely noticeable, especially with the compensation of vocoder.

\begin{table}[htbp]
\footnotesize
\caption{Mean opinion score (MOS) with 95\% CI, real time factor (RTF), and latency (ms) comparison on evaluation set.}
\label{tab:latency_table}
\centering
\begin{tabular}{lccc}
\toprule
Model & MOS & Latency & RTF \\
\midrule
Par. FastPitch    &4.185 $\pm$ 0.043 & 125.77 & 0.029\\
\midrule
Inc. FastPitch (S)&4.178 $\pm$ 0.047 &\multirow{2}{*}{30.35} & \multirow{2}{*}{0.045} \\
Inc. FastPitch (D)&4.145 $\pm$ 0.052 & & \\
\midrule
Ground Truth &4.545 $\pm$ 0.039 & - & -\\
\bottomrule
\end{tabular}
\vspace{0em}
\end{table}

Table 1 also displays RTF and latency. For Incremental FastPitch, RTF is defined as dividing the last chunk's latency by the audio duration, and latency corresponds to the first chunk's latency. The S and D model shares the same inference process. We find that Incremental FastPitch has a higher RTF but is still able to achieve around $22\times$ real-time as it maintains the parallelism of chunk generation. Notably, it has a significantly lower latency compared to parallel FastPitch.

\section{Conclusions}

In this work, we propose Incremental FastPitch, capable of incrementally generating high-quality Mel chunks with low latency while maintaining chunk generation parallelism and consistent computation complexity. We improve the decoder with chunk-based FFT blocks that use fixed size state caching to maintain Mel continuity across chunks. We further experiment with multiple masking configurations of receptive-filed constrained training for adapting model to limited receptive filed inference. Experiments show that our proposal can produce speech quality comparable to the parallel baseline, with a significant lower latency that allows even lower response time for real-time speech synthesis.

\vfill\pagebreak

\bibliographystyle{IEEEbib}
\bibliography{refs}

\end{document}